# Zero-crossing magnetic field modulation at high frequency for triaxial-vectorial SERF atomic magnetometer

J. N. Qin[1,2], X. Zhang[1,2], C. Chen[1,2], Y. Z. Wang[1,2]

[1]Key Laboratory of Geophysical Exploration Equipment, Ministry of Education of China, Jilin University
[2]College of Instrumentation and Electrical Engineering, Jilin University, Changchun, China

The triaxial-vectorial magnetic field measurement method based on the quasi-steady-state solution of Bloch equation for SERF atomic magnetometry is firstly proposed by S. Seltzer and M. Romalis in 2004, which introduces dual-axis zero-crossing magnetic fields modulation[1]. The modulation frequency must be slow enough that the quasi-steady-state solution is valid, otherwise the measurement accuracy will be seriously degraded. Thus the bandwidth of the system is greatly limited. In order to enhance the bandwidth constricted by the range of modulation frequency, we investigate the response of alkali-atomic spin to high-frequency modulation field by solving the Bloch equation in a time-dependence manner. Atomic spin evolution $\boldsymbol{P}$ is described by a conventional Bloch equation[2],

$$\frac{d\boldsymbol{P}}{dt} = \gamma \boldsymbol{P} \times \boldsymbol{B} + R_p(s\hat{z} - \boldsymbol{P}) - R_{rel}\boldsymbol{P} \qquad (1)$$

where $\gamma$ denotes gyromagnetic ratio, $\boldsymbol{B}$ is magnetic field, $R_p$ indicates pumping rate, $s$ indicates the optical-pumping vector along $\hat{z}$ axis, and $R_{rel}$ is total relaxation determined by spin exchange collision, spin destruction collision and so forth.

**1. Modulation in $\hat{x}$ axis.** We introduce a bias magnetic field along $\hat{z}$ axis $B_z = \omega_{z0}/\gamma \hat{z}$ and a sinusoidal field along $\hat{x}$ axis $B_x = B'_x \cos(\omega_x t)\hat{x}$, simultaneously zero the other magnetic components. Assuming that the probing signal is determined by atomic spin along $\hat{x}$ axis $P_x$, then we get the response of equation (1) to the superimposition,

$$P_x = \frac{1}{2} P_0 \gamma B'_x \left[\frac{(\omega_{z0} - \omega_x)}{\Delta\omega^2 + (\omega_{z0} - \omega_x)^2} + \frac{\omega_{z0} + \omega_x}{\Delta\omega^2 + (\omega_{z0} + \omega_x)^2}\right]\cos(\omega_x t)$$
$$+ \frac{1}{2} P_0 \gamma B'_x \left[\frac{\Delta\omega}{\Delta\omega^2 + (\omega_{z0} - \omega_x)^2} - \frac{\Delta\omega}{\Delta\omega^2 + (\omega_{z0} + \omega_x)^2}\right]\sin(\omega_x t) \qquad (2)$$

where $P_0$ is the equilibrium spin polarization and $\Delta\omega$ is the linewidth of magnetic-resonance $\Delta\omega = 2\pi R_{rel}$. Investigating the theoretical analysis (2), the response to various magnetic fields $B_z$ is given by overlapping Lorentzian curves respectively centered at $+\omega_x/\gamma$ and $-\omega_x/\gamma$. Symmetrically, the responsive spectrum to the modulation frequency $\omega_x$ is given by overlapping Lorentzian curves respectively centered at $+\omega_{z0}$ and $-\omega_{z0}$. We numerically compute the formula and obtain the three-dimensional figure, as shown in Fig.1 (a).

**2. Modulation in $\hat{z}$ axis.** We introduce a bias magnetic field along $\hat{x}$ axis $B_x = \omega_{x0}/\gamma \hat{x}$ and a sinusoidal field along $\hat{z}$ axis $B_z = B'_z \sin(\omega_z t)\hat{z}$, simultaneously zero the other magnetic components. We keep the probing signal and all other parameters in consistency with subsection.1, and attain the response of equation (1) to this kind of superimposition,

$$P_x = \frac{1}{2}\gamma B_z' P_0 \frac{\omega_{x0}}{(\omega_{x0}^2 + \Delta\omega^2)} \frac{\Delta\omega}{(\Delta\omega^2 + \omega_z^2)} \cos(\omega_z t) + \frac{1}{2}\gamma B_z' P_0 \frac{\omega_{x0}}{(\omega_{x0}^2 + \Delta\omega^2)} \frac{\omega_z}{(\Delta\omega^2 + \omega_z^2)} \sin(\omega_z t) \quad (3)$$

Investigating the theoretical analysis (3), the response to various $B_x$ is given by a Lorentzian curve centered at 0, whose amplitude varies with modulation frequency $\omega_z$ in a form of Lorentzian shape. Symmetrically, the responsive spectrum to the modulation frequency $\omega_z$ is given by a Lorentzian curve centered at 0, whose amplitude varies with modulation frequency $\omega_{x0}$ in a form of Lorentzian shape, as shown in Fig1. (b).

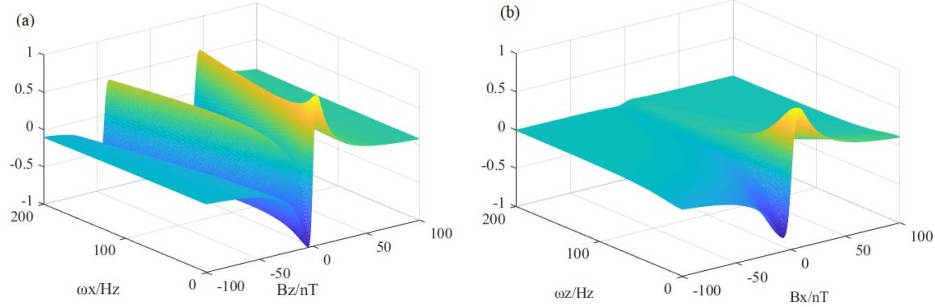

Figure 1：(a) In-phase component of the response calculated by first item of equation (2). (b) In-phase component of the response calculated by first item of equation (3).

## 3. Experiments

We performed a series of experiments to validate the theoretical analyses, as shown in Fig.2 (a) and (b). Fig.2 (a) indicates the response to various magnetic fields $B_z$ under the condition $\omega_x = 33\text{Hz}$ and $\Delta\omega \approx 35\text{Hz}$, Fig.2 (b) indicates the response to various magnetic fields $B_x$ under the condition $\omega_z = 33\text{Hz}, \Delta\omega \approx 35\text{Hz}$. We could find the experimental data agree with the numerical computation. The non-coincident part may caused by non-orthogonality of coils and the inaccurate measurement of $\Delta\omega$.

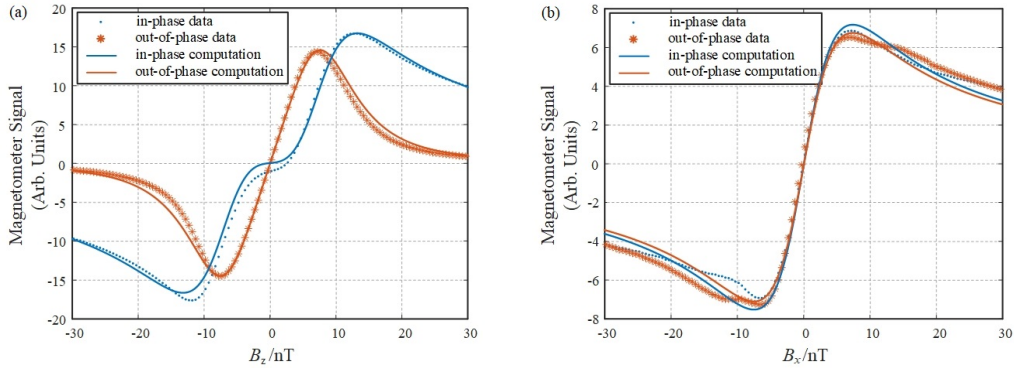

Figure 2：(a) Measurement data and numerical computation of the response to various magnetic fields $B_z$. (b) Measurement data and numerical computation of the response to various magnetic fields $B_x$.

In conclusion, we give the response form of the zero-crossing magnetic modulation at high frequency for triaxial-vectorial SERF atomic magnetometer. The theory can be used to optimize modulation frequency and provide the basis for improving the bandwidth and precision of the closed-loop system.

References
  [1] S Seltzer, M Romalis. Unshielded three-axis vector operation of a spin-exchange-relaxation free atomic magnetometer. Appl. Phys. Lett, 2004, 85(20): 4804-4806.
  [2] S Seltzer. Developments in alkali-mental atomic magnetometry. Princeton: Princeton University, 2008.